\documentclass[english,aps,superscriptaddress]{revtex4}
\usepackage[T1]{fontenc}
\usepackage[latin9]{inputenc}
\usepackage[letterpaper]{geometry}
\usepackage{amsmath}
\usepackage{esint}

\makeatletter
\@ifundefined{textcolor}{}
{%
 \definecolor{BLACK}{gray}{0}
 \definecolor{WHITE}{gray}{1}
 \definecolor{RED}{rgb}{1,0,0}
 \definecolor{GREEN}{rgb}{0,1,0}
 \definecolor{BLUE}{rgb}{0,0,1}
 \definecolor{CYAN}{cmyk}{1,0,0,0}
 \definecolor{MAGENTA}{cmyk}{0,1,0,0}
 \definecolor{YELLOW}{cmyk}{0,0,1,0}
 }

\makeatother

\usepackage{babel}

\begin{document}

\title{A consistent description of kinetic equation with triangle anomaly }

\author{Shi Pu}

\affiliation{Interdisciplinary Center for Theoretical Study and Department of
Modern Physics, University of Science and Technology of China, Hefei
230026, China}

\author{Jian-hua Gao}

\affiliation{Interdisciplinary Center for Theoretical Study and Department of
Modern Physics, University of Science and Technology of China, Hefei
230026, China}

\author{Qun Wang}

\affiliation{Interdisciplinary Center for Theoretical Study and Department of
Modern Physics, University of Science and Technology of China, Hefei
230026, China}
\begin{abstract}
We provide a consistent description of the kinetic equation with triangle
anomaly which is compatible with the entropy principle of the second
law of thermodynamics and the charge/energy-momentum conservation
equations. In general an anomalous source term is necessary to ensure
that the equations for the charge and energy-momentum conservation
are satisfied and that the correction terms of distribution functions
are compatible to these equations. The constraining equations from
the entropy principle are derived for the anomaly-induced leading
order corrections to the particle distribution functions. The correction
terms can be determined for minimum number of unknown coefficients
in one charge and two charge cases by solving the constraining equations. 
\end{abstract}
\maketitle

\section{Introduction}

The experiments at the Relativistic Heavy Ion Collider (RHIC) at Brookhaven
National Laboratory (BNL) provide several pieces of evidence that
the quark gluon plasma (QGP) may have been generated and is strongly
coupled or nearly a perfect fluid (called sQGP), contrary to the conventional
picture of the QGP as a weakly interacting gas of quarks and gluons
(for reviews, see e.g. \cite{Jacobs:2004qv,Shuryak:2004cy,gyulassy:2005}).
Relativistic hydrodynamics \cite{Eckart:1940te,landau:1959,Israel:1976tn,Israel:1979wp}
is a useful tool to describe the space-time evolution of the fireball
formed in heavy ion collisions. The RHIC data for collective flows
have been well described by the ideal and dissipative hydrodynamics
\cite{Huovinen:2001cy,Muronga:2001zk,Kolb:2003dz,Muronga:2004sf,Romatschke:2007mq,Song:2007fn,Baier:2007ix,Betz:2008me}. 

The correspondence of relativistic hydrodynamics to charged black-branes
was investigated by AdS/CFT duality \cite{Erdmenger:2008rm,Banerjee:2008th}.
A new term associated with the axial anomalies was found in the first
order dissipative hydrodynamics (see, e.g., Ref. \cite{Torabian:2009qk}
about holographic hydrodynamics with multiple/non-Abelian symmetries,
or Ref. \cite{Rebhan:2009vc} in Sakai-Sugimoto model). Recently the
new term has been derived in hydrodynamics with a triangle anomaly
\cite{Son:2009tf}. A similar result was also obtained in microscopic
theory of the superfluid \cite{Lublinsky:2009wr}. This problem is
closely related to the so-called Chiral Magnetic Effect (CME) in heavy
ion collisions \cite{Kharzeev:2007jp,Fukushima:2008xe,Fukushima:2010vw,Kharzeev:2010gr}.
When two energetic nuclei pass each other a strong magnetic field
up to $10^{18}$ G is formed, which breaks local parity via axial
anomaly. This effect may be observed through charge separation. Hydrodynamics
in an external background field can be used to pin down the CME in
real time simulation. However the anomalous term in the charge current
breaks the second law of thermodynamics unless new terms of vorticity
and magnetic field are introduced in the charge and entropy currents
\cite{Son:2009tf}. 

In this paper, we try to provide a consistent description of the kinetic
equation with a triangle anomaly. We will derive the kinetic equation
to the next to leading order as well as the leading order correction
to the particle distribution function arising from anomaly. These
results are compatible with the entropy principle of the second law
of thermodynamics and the charge/energy-momentum conservation equations. 

This paper is organized as follows. In Sec. \ref{sec:2}, we will
derive constraining equations from the entropy principle for the correction
terms in distribution functions in the most simple case with one charge
and one particle species (without anti-particles). In Sec. \ref{sec:3},
we will show that an anomaly source term is necessary in general to
ensure that the equations for the charge and energy-momentum conservation
are satisfied and that the correction terms of distribution functions
are compatible to these equations. In Sec. \ref{sec:4} and \ref{sec:5},
we will solve the constraining equations to obtain the correction
terms of distribution functions in the one-charge (with anti-particles)
and two-charges cases. Finally we summarize and make conclusions in
Sec. \ref{sec:6}. We adopt the convention for the metric tensor $g^{\mu\nu}=\mathrm{diag}(+,-,-,-)$.

\section{Constraining distribution function with anomaly compatible to second
law of thermodynamics \label{sec:2}}

In this and the next sections we will consider the most simple case
with one charge and one particle species (without anti-particles).
The relativistic Boltzmann equation for the on-shell phase-space distribution
$f(x,p)$ in a background electromagnetic field $F_{\mu\nu}=\partial_{\mu}A_{\nu}-\partial_{\nu}A_{\mu}$
is given by, 

\begin{equation}
p^{\mu}\left(\frac{\partial}{\partial x^{\mu}}-QF_{\mu\nu}\frac{\partial}{\partial p_{\nu}}\right)f(x,p)=\mathcal{C}[f]\,,\label{boltzmann}\end{equation}
where the charge of the particle is $Q=\pm1$. Here $p$ denotes the
on-shell 4-momentum satisfying $p^{2}=m^{2}$ where $m$ is the particle
mass. We note that $\mathcal{C}[f]$ contains a normal collision term
$\mathcal{C}_{0}[f]$ and a source term from anomaly $\mathcal{C}_{A}[f]$,
$\mathcal{C}[f]=\mathcal{C}_{0}[f]+\mathcal{C}_{A}[f]$. We assume
that $\mathcal{C}_{A}[f]$ is at most of the first order, a small
quantity. The necessity for the source term is to make the charge
conservation equation hold, \begin{equation}
\partial_{\mu}j^{\mu}=-CE^{\mu}B_{\mu}\equiv-CE\cdot B.\label{eq:charge-current}\end{equation}
Here $j^{\mu}$ is the charge current and $E^{\mu}=u_{\nu}F^{\mu\nu}$
and $B_{\mu}=\frac{1}{2}\epsilon_{\mu\nu\alpha\beta}u^{\nu}F^{\alpha\beta}$
are electric and magnetic field vectors respectively, where $u_{\mu}$
is the fluid velocity and $\epsilon_{\mu\nu\alpha\beta}=-\epsilon^{\mu\nu\alpha\beta}=-1,1$
for the order of Lorentz indices $(\mu\nu\alpha\beta)$ is an even/odd
permutation of $(0123)$. However, the presence of the source term
should not influence the energy momentum conservation, \begin{equation}
\partial_{\mu}T^{\mu\nu}=F^{\nu\mu}j_{\mu}.\label{eq:em-cons1}\end{equation}

One can verify that the equilibrium solution of the distribution function,
\begin{equation}
f_{0}=\frac{1}{\exp[\beta u_{\mu}(p^{\mu}-QF^{\mu\nu}x_{\nu})-\beta Q\mu_{0}]-e},\end{equation}
satisfies the collisionless Boltzmann equation (\ref{boltzmann})
in an external field for constant $\beta=1/T$ ($T$ is the local
temperature), $u_{\mu}$ and $\mu_{0}$ (local chemical potential
without electromagnetic field). Here $e=0,\pm1$ for Boltzmann, Bose
and Fermi distributions respectively. When $\beta$, $u_{\mu}$ and
$\mu_{0}$ are not constants but functions of space-time, the Boltzmann
equation (\ref{boltzmann}) is not satisfied automatically. Note that
we can absorb $-Qx_{\nu}u_{\mu}F^{\mu\nu}=Qx\cdot E$ into $\mu_{0}$
so that $f_{0}$ has the form of an equilibrium distribution function, 

\begin{eqnarray}
f_{0}(x,p) & = & \frac{1}{e^{(u\cdot p-Q\mu)/T}-e},\label{eq:free-f}\end{eqnarray}
where $\mu\equiv\mu_{0}-x\cdot E$. 

We assume that the distribution function $f$ in presence of an anomaly
is a solution of the Boltzmann equation with collision terms in Eq.
(\ref{boltzmann}), where $\beta$, $u_{\mu}$ and $\mu$ are functions
of space-time. Generally $f(x,p)$ can be written in the following
form, 

\begin{eqnarray}
f(x,p) & = & \frac{1}{e^{(u\cdot p-Q\mu)/T+\chi(x,p)}-e}=f_{0}(x,p)+f_{1}(x,p),\label{eq:fxp-expansion}\end{eqnarray}
where $f_{0}(x,p)$ is given in Eq. (\ref{eq:free-f}) and $f_{1}(x,p)$
is the first order deviation from it, 

\begin{eqnarray}
f_{1}(x,p) & = & -f_{0}(x,p)\left[1+ef_{0}(x,p)\right]\chi(x,p).\label{f1xp}\end{eqnarray}
It is known that a magnetic field is closely related to a charge rotation
characterized by vorticity. So we introduce into the distribution
function terms associated with the vorticity-induced current $\omega_{\mu}=\frac{1}{2}\epsilon_{\mu\nu\alpha\beta}u^{\nu}\partial^{\alpha}u^{\beta}$
and the magnetic field 4-vector $B_{\mu}$ which are assumed to be
of the first order, which provide a leading order correction to the
particle distribution function. For simplicity we will neglect viscous
and diffusive effects throughout the paper, then the ordinary form
in the current scheme for $\chi(x,p)$ reads, 

\begin{eqnarray}
\chi(x,p) & = & \lambda(p)p\cdot\omega+\lambda_{B}(p)p\cdot B,\label{eq:deviation}\end{eqnarray}
where $\lambda(p)$ and $\lambda_{B}(p)$ are functions of $\mu$,
$T$, and $u\cdot p$ and have mass dimension $-2$ and $-3$ respectively.
We will show that $\lambda(p)$ and $\lambda_{B}(p)$ must depend
on momentum otherwise they will contradict the entropy principle from
the second law of thermodynamics. 

Using Eq. (\ref{eq:fxp-expansion}) we can decompose the charge and
entropy currents and the stress tensor into equilibrium values and
the leading order (first order) corrections as $j^{\mu}=j_{0}^{\mu}+j_{1}^{\mu}$,
$S^{\mu}=S_{0}^{\mu}+S_{1}^{\mu}$ and $T^{\mu\nu}=T_{0}^{\mu\nu}+T_{1}^{\mu\nu}$
with \begin{eqnarray}
j_{0,1}^{\mu}(x) & = & Q\int[dp]p^{\mu}f_{0,1}(x,p),\nonumber \\
S_{0}^{\mu}(x) & = & -\int[dp]p^{\mu}\psi(f_{0}),\nonumber \\
S_{1}^{\mu}(x) & = & -\int[dp]p^{\mu}\psi'(f_{0})f_{1},\nonumber \\
T_{0,1}^{\mu\nu}(x) & = & \int[dp]p^{\mu}p^{\nu}f_{0,1}(x,p),\label{eq:j1s1t1}\end{eqnarray}
where we have defined $[dp]\equiv d_{g}\frac{d^{3}p}{(2\pi)^{3}(u\cdot p)}$
($d_{g}$ is the degeneracy factor), $\psi(f_{0})=f_{0}\ln(f_{0})-e(1+ef_{0})\ln(1+ef_{0})$
and $\psi'(f_{0})=\ln[f_{0}/(1+ef_{0})]=-(u\cdot p-Q\mu)/T$. Inserting
$f_{0}$ into the above formula, we obtain the charge and entropy
currents and the stress tensor in equilibrium, $j_{0}^{\mu}=nu^{\mu}$,
$S_{0}^{\mu}=su^{\mu}$ and $T_{0}^{\mu\nu}=(\epsilon+P)u^{\mu}u^{\nu}-Pg^{\mu\nu}$,
with the energy density $\varepsilon$, the pressure $P$, the particle
number density $n$ and the entropy density $s=(\epsilon+P-n\mu)/T$.
Using Eqs. (\ref{f1xp},\ref{eq:deviation},\ref{eq:j1s1t1}), we
obtain 

\begin{eqnarray}
j_{1}^{\mu} & = & \xi\omega^{\mu}+\xi_{B}B^{\mu},\nonumber \\
T_{1}^{\mu\nu} & = & DT\left(u^{\mu}\omega^{\nu}+u^{\nu}\omega^{\mu}\right)+D_{B}T\left(u^{\mu}B^{\nu}+u^{\nu}B^{\mu}\right),\nonumber \\
S_{1}^{\mu} & = & -\frac{\mu}{T}(\xi\omega^{\mu}+\xi_{B}B^{\mu})+(D\omega^{\mu}+D_{B}B^{\mu}),\label{eq:j-t-s1}\end{eqnarray}
where \begin{eqnarray}
\xi & = & -QJ_{21}^{\lambda}\equiv\frac{1}{3}Q\int[dp][(p\cdot u)^{2}-m^{2}]f_{0}(1+ef_{0})\lambda(p),\nonumber \\
\xi_{B} & = & -QJ_{21}^{\lambda_{B}}\equiv\frac{1}{3}Q\int[dp][(p\cdot u)^{2}-m^{2}]f_{0}(1+ef_{0})\lambda_{B}(p),\nonumber \\
D & = & -\frac{J_{31}^{\lambda}}{T}\equiv\frac{1}{3T}\int[dp][(p\cdot u)^{2}-m^{2}](p\cdot u)f_{0}(1+ef_{0})\lambda(p),\nonumber \\
D_{B} & = & -\frac{J_{31}^{\lambda_{B}}}{T}\equiv\frac{1}{3T}\int[dp][(p\cdot u)^{2}-m^{2}](p\cdot u)f_{0}(1+ef_{0})\lambda_{B}(p).\label{eq:db}\end{eqnarray}

On the other hand, $\xi$, $\xi_{B}$, $D$ and $D_{B}$ as functions
of $\mu$ and $T$ can be determined by the second law of thermodynamics
or the entropy principle together with Eq. (\ref{eq:charge-current}-\ref{eq:em-cons1}).
We find that $\partial_{\mu}(su^{\mu}+S_{1}^{\mu})$ cannot be positive
definite unless we make a shift to introduce a new entropy current
$\tilde{S}^{\mu}$ as follows, \begin{eqnarray}
\tilde{S}^{\mu} & = & su^{\mu}+S_{1}^{\mu}-(D\omega^{\mu}+D_{B}B^{\mu})=su^{\mu}-\frac{\mu}{T}\left(\xi\omega^{\mu}+\xi_{B}B^{\mu}\right)\nonumber \\
 & = & \frac{1}{T}(Pu^{\mu}-\mu j^{\mu}+u_{\lambda}T^{\lambda\mu})-(D\omega^{\mu}+D_{B}B^{\mu})\end{eqnarray}
We will use the thermodynamic relation \begin{equation}
\partial_{\mu}(Pu^{\mu})=j_{0}^{\mu}\partial_{\mu}\overline{\mu}-T_{0}^{\lambda\mu}\partial_{\mu}u_{\lambda}\label{eq:d-pu}\end{equation}
and the identities \begin{eqnarray}
u^{\mu}u^{\lambda}\partial_{\mu}\omega_{\lambda} & = & \frac{1}{2}\partial_{\mu}\omega^{\mu},\nonumber \\
u^{\mu}u^{\lambda}\partial_{\mu}B_{\lambda} & = & \partial_{\mu}B^{\mu}-2\omega^{\rho}E_{\rho},\nonumber \\
\partial_{\mu}\omega^{\mu} & = & -\frac{2}{\epsilon+P}(n\omega^{\mu}E_{\mu}+\omega^{\mu}\partial_{\mu}P),\nonumber \\
\partial_{\mu}B^{\mu} & = & 2\omega^{\rho}E_{\rho}-\frac{1}{\epsilon+P}(nB_{\lambda}E^{\lambda}+B^{\mu}\partial_{\mu}P),\label{eq:id-oeb}\end{eqnarray}
to evaluate $\partial_{\mu}\tilde{S}^{\mu}$. We have used the shorthand
notation $\overline{\mu}\equiv\mu/T$ in Eq. (\ref{eq:d-pu}). Following
the same procedure as in Ref. \cite{Son:2009tf}, we obtain \begin{eqnarray}
\partial_{\mu}\widetilde{S}^{\mu} & = & \omega^{\mu}\left[\xi^{SS}\partial_{\mu}\overline{\mu}-\partial_{\mu}D+\frac{2D}{\epsilon+P}\partial_{\mu}P\right]\nonumber \\
 &  & +B^{\mu}\left[\xi_{B}^{SS}\partial_{\mu}\overline{\mu}-\partial_{\mu}D_{B}+\frac{D_{B}}{\epsilon+P}\partial_{\mu}P\right]\nonumber \\
 &  & +E\cdot\omega\left[\frac{1}{T}\xi^{SS}+\frac{2nD}{\epsilon+P}-2D_{B}\right]\nonumber \\
 &  & +E\cdot B\left[\frac{1}{T}\xi_{B}^{SS}+C\frac{\mu_{A}}{T}+\frac{nD_{B}}{\epsilon+P}\right].\label{eq:ds-tilde}\end{eqnarray}
where we have defined \begin{eqnarray}
\xi^{SS} & = & \frac{DTn}{\epsilon+P}-\xi,\;\xi_{B}^{SS}=\frac{D_{B}Tn}{\epsilon+P}-\xi_{B}.\label{eq:ss-xi}\end{eqnarray}
For the constraint $\partial_{\mu}\widetilde{S}^{\mu}\geq0$ to hold,
we impose that all quantities inside the square brackets should vanish.
We finally obtain \begin{equation}
D=\frac{1}{3}C\frac{\mu^{3}}{T},\; D_{B}=\frac{1}{2}C\frac{\mu^{2}}{T},\;\xi=-C\frac{sT\mu^{2}}{\epsilon+P},\;\xi_{B}=-C\frac{sT\mu}{\epsilon+P}.\label{eq:entropy-d}\end{equation}
Using Eqs. (\ref{eq:ss-xi},\ref{eq:entropy-d}), one can verify that
the values of $\xi^{SS}$ and $\xi_{B}^{SS}$ are identical to Ref.
\cite{Son:2009tf}. The difference between our values in Eq. (\ref{eq:entropy-d})
and those in Ref. \cite{Son:2009tf} arises from the fact that we
do not use the Landau frame while the authors of Ref. \cite{Son:2009tf}
do. By equating Eq. (\ref{eq:db}) and (\ref{eq:entropy-d}), we obtain
equations for $\lambda$ and $\lambda_{B}$, \begin{eqnarray}
 &  & QJ_{21}^{\lambda}=-\xi,\; J_{31}^{\lambda}=-DT,\; QJ_{21}^{\lambda_{B}}=-\xi_{B},\; J_{31}^{\lambda_{B}}=-D_{B}T.\label{eq:constraint-lambda}\end{eqnarray}
Equation (\ref{eq:constraint-lambda}) forms a complete set of constraints
for $\lambda$ and $\lambda_{B}$. We note that $\lambda$ and $\lambda_{B}$
must depend on momentum in general. If $\lambda$ and $\lambda_{B}$
are constants, we would obtain \begin{equation}
\frac{\xi}{DT}=\frac{\xi_{B}}{D_{B}T}=Q\frac{J_{21}}{J_{31}},\end{equation}
which contradict Eq. (\ref{eq:entropy-d}) from the entropy principle. 

We can expand $\lambda(p)$ and $\lambda_{B}(p)$ in powers of $u\cdot p$,
\begin{equation}
\lambda(p)=\sum_{i=0}\lambda_{i}(u\cdot p)^{i},\;\lambda_{B}(p)=\sum_{i=0}\lambda_{i}^{B}(u\cdot p)^{i}.\label{eq:lambda-expand}\end{equation}
So we obtain the following expressions \begin{equation}
J_{n1}^{\lambda}=\sum_{i=0}\lambda_{i}J_{i+n,1},\; J_{n1}^{\lambda_{B}}=\sum_{i=0}\lambda_{i}^{B}J_{i+n,1},\label{eq:j-expand}\end{equation}
for $n=2,3$. Here the functions $J_{nq}$ are integrals defined in
Ref. \cite{Israel:1979wp,Muronga:2001zk}, \begin{equation}
J_{nq}=(-1)^{q}\frac{1}{(2q+1)!!}\int\frac{d^{3}p}{(2\pi)^{3}(u\cdot p)}[(u\cdot p)^{2}-m^{2}]^{q}(u\cdot p)^{n-2q}f_{0}(1+ef_{0}),\label{eq:jnk}\end{equation}
Using Eqs. (\ref{eq:lambda-expand},\ref{eq:j-expand}) in Eq. (\ref{eq:constraint-lambda}),
we can constrain the coefficients $\lambda_{i}$ and $\lambda_{i}^{B}$.
If we expand both $\lambda(p)$ and $\lambda_{B}(p)$ to the first
power of $u\cdot p$, we can completely fix the coefficients $\lambda_{0,1}$
and $\lambda_{0,1}^{B}$ from Eq. (\ref{eq:constraint-lambda}) since
we have two equations for $\lambda_{0,1}$ and two for $\lambda_{0,1}^{B}$,
\begin{eqnarray}
\left(\begin{array}{cc}
QJ_{21} & QJ_{31}\\
J_{31} & J_{41}\end{array}\right)\left(\begin{array}{c}
\lambda_{0}\\
\lambda_{1}\end{array}\right) & = & \left(\begin{array}{c}
-\xi\\
-DT\end{array}\right),\label{eq:lambda01-1p}\end{eqnarray}
whose solutions to $\lambda_{0,1}$ are \begin{eqnarray}
\left(\begin{array}{c}
\lambda_{0}\\
\lambda_{1}\end{array}\right) & = & \frac{1}{Q(J_{21}J_{41}-J_{31}^{2})}\left(\begin{array}{c}
-\xi J_{41}+DTQJ_{31}\\
\xi J_{31}-DTQJ_{21}\end{array}\right).\label{eq:sol-lambda01-1p}\end{eqnarray}
The equations and solutions for $\lambda_{0,1}^{B}$ are in the same
form as Eqs. (\ref{eq:lambda01-1p},\ref{eq:sol-lambda01-1p}) with
replacements $\lambda_{0,1}\rightarrow\lambda_{0,1}^{B}$, $\xi\rightarrow\xi_{B}$
and $D\rightarrow D_{B}$. 

For massless fermions and small $\overline{\mu}$, the results are
\begin{equation}
\lambda_{0}\approx-CG_{1}\frac{\mu^{2}}{T^{4}},\;\lambda_{1}\approx CG_{2}\frac{\mu^{2}}{T^{5}},\;\lambda_{0}^{B}\approx\frac{\lambda_{0}}{\mu},\;\lambda_{1}^{B}\approx\frac{\lambda_{1}}{\mu},\end{equation}
where $G_{1}$ and $G_{2}$ are two constants, $G_{1}\equiv607500\pi^{2}\zeta(5)/(d_{g}G_{0})$
and $G_{2}\equiv1260\pi^{6}/(d_{g}G_{0})$ with $G_{0}\equiv455625\zeta(3)\zeta(5)-49\pi^{8}$.
We notice that the $D$ terms in Eq. (\ref{eq:sol-lambda01-1p}) are
negligible, so the solutions are proportional to $\xi$.

\section{Collision and anomalous source terms \label{sec:3}}

In this section we will show that a general form of $\lambda(p)$
and $\lambda_{B}(p)$ are compatible to the charge and energy-momentum
conservation equations (\ref{eq:charge-current}) and (\ref{eq:em-cons1}).
We will also derive equations for the collision and anomalous source
terms. For simplicity we consider the single charge case without anti-particles. 

The charge conservation equation (\ref{eq:charge-current}) can be
derived from the Boltzmann equation (\ref{boltzmann}) as \begin{eqnarray}
\partial_{\mu}j^{\mu}(x) & = & \int\frac{d^{3}p}{(2\pi)^{3}E_{p}}p^{\mu}\partial_{\mu}f(x,p)=\int\frac{d^{3}p}{(2\pi)^{3}E_{p}}p^{\mu}F_{\mu\nu}\frac{\partial f}{\partial p_{\nu}}+\int\frac{d^{3}p}{(2\pi)^{3}E_{p}}\mathcal{C}[f]\nonumber \\
 & = & \int\frac{d^{3}p}{(2\pi)^{3}E_{p}}\mathcal{C}[f],\label{eq:djx}\end{eqnarray}
where have used the identity \begin{eqnarray}
\int\frac{d^{3}p}{2E_{p}}p^{\mu}F_{\mu\nu}\frac{\partial f}{\partial p_{\nu}} & = & \int d^{4}p\theta(p_{0})\delta(p^{2}-m^{2})p^{\mu}F_{\mu\nu}\frac{\partial f}{\partial p_{\nu}}\nonumber \\
 & = & -\int d^{4}p\frac{\partial}{\partial p_{\nu}}[\theta(p_{0})\delta(p^{2}-m^{2})p^{\mu}F_{\mu\nu}]f=0.\label{eq:property-1}\end{eqnarray}
Then the momentum integral of the collision term must obey \begin{equation}
\int\frac{d^{3}p}{(2\pi)^{3}E_{p}}\mathcal{C}[f]=-CE\cdot B,\label{eq:constraint1}\end{equation}
so that Eq. (\ref{eq:charge-current}) can hold. The energy and momentum
conservation equation (\ref{eq:em-cons1}) can be derived from the
Boltzmann equation (\ref{boltzmann}) as \begin{eqnarray}
\partial_{\mu}T^{\mu\nu} & = & \int\frac{d^{3}p}{(2\pi)^{3}E_{p}}p^{\mu}p^{\nu}\partial_{\mu}f=\int\frac{d^{3}p}{(2\pi)^{3}E_{p}}p^{\nu}p^{\alpha}F_{\alpha\beta}\frac{\partial f}{\partial p_{\beta}}+\int\frac{d^{3}p}{(2\pi)^{3}E_{p}}p^{\nu}\mathcal{C}[f]\nonumber \\
 & = & F^{\nu\mu}j_{\mu}+\int\frac{d^{3}p}{(2\pi)^{3}E_{p}}p^{\nu}\mathcal{C}[f],\end{eqnarray}
where we have used \begin{eqnarray}
\int\frac{d^{3}p}{2E_{p}}p^{\nu}p^{\alpha}F_{\alpha\beta}\frac{\partial f}{\partial p_{\beta}} & = & \int d^{4}p\theta(p_{0})\delta(p^{2}-m^{2})p^{\nu}p^{\alpha}F_{\alpha\beta}\frac{\partial f}{\partial p_{\beta}}\nonumber \\
 & = & -\int d^{4}p\frac{\partial}{\partial p_{\beta}}[\theta(p_{0})\delta(p^{2}-m^{2})p^{\nu}p^{\alpha}F_{\alpha\beta}]f=F^{\nu\alpha}j_{\alpha}.\label{eq:property-2}\end{eqnarray}
Then we require that the collision term must satisfy \begin{equation}
\int\frac{d^{3}p}{(2\pi)^{3}E_{p}}p^{\nu}\mathcal{C}[f]=0,\label{eq:constraint2}\end{equation}
so that Eq. (\ref{eq:em-cons1}) can hold. 

We can expand $\mathcal{C}[f]$ to the second order as \begin{eqnarray}
\mathcal{C}[f] & = & \mathcal{C}_{0}[f_{0}+f_{1}+f_{2}]+\mathcal{C}_{A}[f]\approx\mathcal{C}_{1}+\mathcal{C}_{2}\label{eq:collision-expansion}\end{eqnarray}
where we have used the property $\mathcal{C}_{0}[f_{0}]=0$, and defined
\begin{eqnarray}
\mathcal{C}_{1} & = & \left.\frac{d\mathcal{C}_{0}}{df}\right|_{f=f_{0}}f_{1}+\mathcal{C}_{A1},\nonumber \\
\mathcal{C}_{2} & = & \left.\frac{d\mathcal{C}_{0}}{df}\right|_{f=f_{0}}f_{2}+\frac{1}{2}\left.\frac{d^{2}\mathcal{C}_{0}}{df^{2}}\right|_{f=f_{0}}f_{1}^{2}+\mathcal{C}_{A2}.\label{eq:c1c2}\end{eqnarray}
Note that the general form for the normal part of $\mathcal{C}_{1}$
is $\left.\frac{d\mathcal{C}_{0}}{df}\right|_{f=f_{0}}f_{1}=H_{\lambda}(u\cdot p)p\cdot\omega+H_{\lambda_{B}}(u\cdot p)p\cdot B$.
When inserting the distribution function (\ref{eq:fxp-expansion})
into the Boltzmann equation (\ref{boltzmann}) and using Eq. (\ref{eq:collision-expansion}),
we obtain the Boltzmann equations to the first and second order, \begin{eqnarray}
p^{\mu}\left(\frac{\partial}{\partial x^{\mu}}-F_{\mu\nu}\frac{\partial}{\partial p_{\nu}}\right)f_{0} & = & \mathcal{C}_{1},\label{eq:first-order}\\
p^{\mu}\left(\frac{\partial}{\partial x^{\mu}}-F_{\mu\nu}\frac{\partial}{\partial p_{\nu}}\right)f_{1} & = & \mathcal{C}_{2}.\label{eq:second-order}\end{eqnarray}
From Eqs. (\ref{eq:c1c2},\ref{eq:first-order}) we can determine
the anomalous source term of the first order, \begin{equation}
\mathcal{C}_{A1}=-H_{\lambda}(u\cdot p)p\cdot\omega-H_{\lambda_{B}}(u\cdot p)p\cdot B-f_{0}(1+ef_{0})p^{\mu}\partial_{\mu}[(u\cdot p-\mu_{0})/T].\label{eq:ca1}\end{equation}
By evaluating the left hand sides of Eq. (\ref{eq:second-order}),
we can fix $\mathcal{C}_{2}$ as follows \begin{eqnarray}
\mathcal{C}_{2} & = & f_{0}(1+ef_{0})\left\{ -[(1+2ef_{0})\lambda\partial_{\mu}\psi'(f_{0})+(\partial_{\mu}\lambda)]p^{\mu}p^{\nu}\omega_{\nu}\right.\nonumber \\
 &  & -[(1+2ef_{0})\lambda_{B}\partial_{\mu}\psi'(f_{0})+(\partial_{\mu}\lambda_{B})]p^{\mu}p^{\nu}B_{\nu}\nonumber \\
 &  & +\left[-\beta(1+2ef_{0})\lambda+\frac{d\lambda}{d(u\cdot p)}\right]p^{\mu}p^{\nu}E_{\mu}\omega_{\nu}+\lambda p^{\mu}F_{\mu\nu}\omega^{\nu}\nonumber \\
 &  & +\left[-\beta(1+2ef_{0})\lambda_{B}+\frac{d\lambda_{B}}{d(u\cdot p)}\right]p^{\mu}p^{\nu}E_{\mu}B_{\nu}+\lambda_{B}p^{\mu}F_{\mu\nu}B^{\nu}\nonumber \\
 &  & \left.-\lambda p^{\mu}p^{\nu}\partial_{\mu}\omega_{\nu}-\lambda_{B}p^{\mu}p^{\nu}\partial_{\mu}B_{\nu}\right\} \label{eq:second-kinetic-2}\end{eqnarray}

Taking momentum integrals for Eqs. (\ref{eq:first-order},\ref{eq:second-order}),
we obtain the divergences of charge currents to the first and second
order, \begin{eqnarray}
\partial_{\mu}j_{0}^{\mu}(x) & = & \int[dp]\mathcal{C}_{1}=\int[dp]\mathcal{C}_{A1},\label{eq:id-1st}\\
\partial_{\mu}j_{1}^{\mu}(x) & = & \int[dp]\mathcal{C}_{2}.\label{eq:id-2nd}\end{eqnarray}
where we have used the property, \begin{equation}
\int[dp]\left.\frac{d\mathcal{C}_{0}}{df}\right|_{f=f_{0}}f_{1}=\int[dp][H_{\lambda}(u\cdot p)p\cdot\omega+H_{\lambda_{B}}(u\cdot p)p\cdot B]=0.\end{equation}
With Eq. (\ref{eq:j1s1t1}) and identities in Eq. (\ref{eq:id-oeb}),
the left hand side of Eq. (\ref{eq:id-2nd}) can be evaluated as,
\begin{eqnarray}
\partial_{\mu}j_{1}^{\mu}(x) & = & \omega\cdot E\left(2\xi_{B}-\frac{2n\xi}{\epsilon+P}\right)-B\cdot E\frac{n\xi_{B}}{\epsilon+P}\nonumber \\
 &  & +\omega^{\mu}\left(\partial_{\mu}\xi-\frac{2\xi}{\epsilon+P}\partial_{\mu}P\right)+B^{\mu}\left(\partial_{\mu}\xi_{B}-\frac{\xi_{B}}{\epsilon+P}\partial_{\mu}P\right).\label{eq:lhs-j1}\end{eqnarray}
In order for Eq. (\ref{eq:charge-current}) to be satisfied, it is
required that $\partial_{\mu}j_{0}^{\mu}(x)$ must have the form \begin{eqnarray}
\partial_{\mu}j_{0}^{\mu}(x) & = & -CB\cdot E-\partial_{\mu}j_{1}^{\mu}(x)\nonumber \\
 & = & -\omega\cdot E\left(2\xi_{B}-\frac{2n\xi}{\epsilon+P}\right)-B\cdot E\left(C-\frac{n\xi_{B}}{\epsilon+P}\right)\nonumber \\
 &  & -\omega^{\mu}\left(\partial_{\mu}\xi-\frac{2\xi}{\epsilon+P}\partial_{\mu}P\right)-B^{\mu}\left(\partial_{\mu}\xi_{B}-\frac{\xi_{B}}{\epsilon+P}\partial_{\mu}P\right).\label{eq:lhs-j0}\end{eqnarray}
We see that $\partial_{\mu}j_{0}^{\mu}(x)$ is not vanishing but of
the second order though it is superficially a first order quantity.
This is very important otherwise it would lead to $\partial_{\mu}j_{1}^{\mu}(x)=-CE\cdot B$
and give rise to incompatible results for $\xi,\xi_{B},D,D_{B}$ with
Ref. \cite{Son:2009tf}. Therefore we should keep in mind that it
is the full charge current that satisfies charge conservation equation
(\ref{eq:charge-current}). We note that the expression of $\partial_{\mu}j_{0}^{\mu}(x)$
in Eq. (\ref{eq:lhs-j0}) also gives the momentum integral of the
anomalous source term of the first order, \begin{eqnarray}
\int[dp]\mathcal{C}_{A1} & = & -\omega\cdot E\left(2\xi_{B}-\frac{2n\xi}{\epsilon+P}\right)-B\cdot E\left(C-\frac{n\xi_{B}}{\epsilon+P}\right)\nonumber \\
 &  & -\omega^{\mu}\left(\partial_{\mu}\xi-\frac{2\xi}{\epsilon+P}\partial_{\mu}P\right)-B^{\mu}\left(\partial_{\mu}\xi_{B}-\frac{\xi_{B}}{\epsilon+P}\partial_{\mu}P\right).\label{eq:ca1-int}\end{eqnarray}

For the energy and momentum conservation we obtain, \begin{eqnarray}
\partial_{\mu}T_{0}^{\mu\nu}-F^{\nu\mu}j_{\mu}^{0} & = & \int[dp]p^{\nu}\mathcal{C}_{1},\label{eq:en-1st}\\
\partial_{\mu}T_{1}^{\mu\nu}-F^{\nu\mu}j_{\mu}^{1} & = & \int[dp]p^{\nu}\mathcal{C}_{2},\label{eq:en-2nd}\end{eqnarray}
With Eqs. (\ref{eq:j1s1t1},\ref{eq:id-oeb}), we can evaluate the
left hand side of Eq. (\ref{eq:en-2nd}) as, \begin{eqnarray}
\partial_{\mu}T_{1}^{\mu\nu}-F^{\nu\mu}j_{\mu}^{1} & = & \omega^{\nu}[DT\partial\cdot u+u\cdot\partial(DT)]+B^{\nu}[D_{B}T\partial\cdot u+u\cdot\partial(D_{B}T)]\nonumber \\
 &  & +u^{\nu}\{\omega\cdot E(2D_{B}T-\frac{2DTn}{\epsilon+P})-E\cdot B\frac{D_{B}Tn}{\epsilon+P}+\omega\cdot[\partial(DT)-\frac{2DT}{\epsilon+P}\partial P]\nonumber \\
 &  & +B\cdot[\partial(D_{B}T)-\frac{D_{B}T}{\epsilon+P}\partial P]\}\nonumber \\
 &  & +DT(u\cdot\partial\omega^{\nu}+\omega\cdot\partial u^{\nu})+D_{B}T(u\cdot\partial B^{\nu}+B\cdot\partial u^{\nu})\nonumber \\
 &  & -F^{\nu\mu}(\xi\omega_{\mu}+\xi_{B}B_{\mu}),\label{eq:en-ob1}\end{eqnarray}
where one can verify that each term in the right hand side is of second
order. The left hand side of Eq. (\ref{eq:en-1st}) is then given
by \begin{equation}
\partial_{\mu}T_{0}^{\mu\nu}-F^{\nu\mu}j_{\mu}^{0}=-(\partial_{\mu}T_{1}^{\mu\nu}-F^{\nu\mu}j_{\mu}^{1}),\label{eq:en-ob0}\end{equation}
which is also a second order quantity though $p^{\nu}\mathcal{C}_{1}=p^{\nu}p^{\mu}\left(\frac{\partial}{\partial x^{\mu}}-F_{\mu\sigma}\frac{\partial}{\partial p_{\sigma}}\right)f_{0}$
is superficially of first order. One might question the validity of
Eq. (\ref{eq:id-oeb}) which follows $\partial_{\mu}T_{0}^{\mu\nu}-F^{\nu\mu}j_{\mu}^{0}=0$,
but it is not a problem here since this equation really holds at the
first order or the leading order but not true at the second order.
It is essential that the energy momentum equation (\ref{eq:em-cons1})
hold for the full quantities $T^{\mu\nu}$ and $j_{\mu}$, and not
for $T_{0,1}^{\mu\nu}$ and $j_{\mu}^{0,1}$ separately, otherwise
the results would be contradictory to those of Ref. \cite{Son:2009tf}
following the entropy principle of the second law of thermodynamics. 

We now obtain the momentum integral of $p^{\nu}\mathcal{C}_{1,2}$
from Eqs. (\ref{eq:en-1st},\ref{eq:en-2nd}) with $\partial_{\mu}T_{0,1}^{\mu\nu}-F^{\nu\mu}j_{\mu}^{0,1}$
given by Eqs.(\ref{eq:en-ob1},\ref{eq:en-ob0}).

\section{One charge with particle/anti-particle \label{sec:4}}

We now consider one charge case but add anti-particles to the system.
We will calculate $\lambda(p)$ and $\lambda_{B}(p)$. Since there
are particles and anti-particles, we recover the index $Q=\pm1$ in
particle distribution function, $f\rightarrow f^{Q}$ and $f_{0,1}\rightarrow f_{0,1}^{Q}$.
In Eq. (\ref{eq:j1s1t1}), summations over $Q$ should be added. In
Eq. (\ref{eq:db}) we also have to add the index $Q$ to $J_{n1}^{\lambda}$
and $J_{n1}^{\lambda_{B}}$: $J_{n1}^{\lambda}\rightarrow J_{n1}^{\lambda,Q}$
and $J_{n1}^{\lambda_{B}}\rightarrow J_{n1}^{\lambda_{B},Q}$ ($n=2,3$),
and add summations over $Q$ into the formula of $\xi$, $\xi_{B}$,
$D$, $D_{B}$. Inserting $f_{0}^{Q}$ into Eq. (\ref{eq:j1s1t1}),
we obtain the charge and entropy currents and the stress tensor in
equilibrium, $j_{0}^{\mu}=nu^{\mu}$, $S_{0}^{\mu}=su^{\mu}$ and
$T_{0}^{\mu\nu}=(\epsilon+P)u^{\mu}u^{\nu}-Pg^{\mu\nu}$, with the
particle number density $n\equiv\sum_{Q}Qn_{Q}$, the energy density
$\epsilon=\sum_{Q}\epsilon_{Q}$, the pressure $P=\sum_{Q}P_{Q}$,
and the entropy density $s=\sum_{Q}s_{Q}=(\epsilon+P-n\mu)/T$. The
solutions to $\xi,\xi_{B},D,D_{B}$ are the same as in Eq. (\ref{eq:entropy-d}).
Then Eq. (\ref{eq:constraint-lambda}) is modified to \begin{eqnarray}
 &  & J_{21}^{\lambda,+}-J_{21}^{\lambda,-}=-\xi,\; J_{31}^{\lambda,+}+J_{31}^{\lambda,+}=-DT,\nonumber \\
 &  & J_{21}^{\lambda_{B},+}-J_{21}^{\lambda_{B},-}=-\xi_{B},\; J_{31}^{\lambda_{B},+}+J_{31}^{\lambda_{B},-}=-D_{B}T.\end{eqnarray}
Eq. (\ref{eq:j-expand}) now becomes \begin{equation}
J_{n1}^{\lambda,Q}=\sum_{i=0}\lambda_{i}J_{i+n,1}^{Q},\; J_{n1}^{\lambda_{B},Q}=\sum_{i=0}\lambda_{i}^{B}J_{i+n,1}^{Q},\end{equation}
In the case of minimal number of coefficients we can completely fix
the coefficients $\lambda_{0,1}$ by solving following system of equations,
\begin{eqnarray}
\left(\begin{array}{cc}
\delta J_{21} & \delta J_{31}\\
\sigma J_{31} & \sigma J_{41}\end{array}\right)\left(\begin{array}{c}
\lambda_{0}\\
\lambda_{1}\end{array}\right) & = & \left(\begin{array}{c}
-\xi\\
-DT\end{array}\right),\label{eq:lambda01}\end{eqnarray}
where we have used the shorthand notation, $\delta J_{n1}\equiv J_{n1}^{+}-J_{n1}^{-}$
and $\sigma J_{n1}\equiv J_{n1}^{+}+J_{n1}^{-}$. The solutions to
$\lambda_{0,1}$ are \begin{eqnarray}
\left(\begin{array}{c}
\lambda_{0}\\
\lambda_{1}\end{array}\right) & = & \frac{1}{\Delta}\left(\begin{array}{c}
-\xi(\sigma J_{41})+DT(\delta J_{31})\\
\xi(\sigma J_{31})-DT(\delta J_{21})\end{array}\right),\label{eq:sol-lambda01}\end{eqnarray}
where $\Delta=(\delta J_{21})(\sigma J_{41})-(\sigma J_{31})(\delta J_{31})$.
The equations and solutions for $\lambda_{0,1}^{B}$ are in the same
form as Eqs. (\ref{eq:lambda01},\ref{eq:sol-lambda01}) with replacements
$\lambda_{0,1}\rightarrow\lambda_{0,1}^{B}$, $\xi\rightarrow\xi_{B}$
and $D\rightarrow D_{B}$. 

For massless fermions and small $\overline{\mu}$, we have, \begin{eqnarray}
\sigma J_{21} & \approx & 9\zeta(3)G,\;\sigma J_{31}\approx\frac{7\pi^{4}}{15}GT,\;\sigma J_{41}\approx225\zeta(5)GT^{2},\nonumber \\
\delta J_{21} & \approx & \pi^{2}\overline{\mu}G,\;\delta J_{31}\approx36\zeta(3)\overline{\mu}GT,\;\delta J_{41}\approx\frac{7\pi^{4}}{3}\overline{\mu}GT^{2},\nonumber \\
\Delta & \approx & \frac{\pi^{2}}{5}G^{2}G_{0}T^{2}\overline{\mu}.\end{eqnarray}
where $G\equiv\frac{d_{g}T^{4}}{6\pi^{2}}$ and $G_{0}\equiv-84\pi^{2}\zeta(3)+1125\zeta(5)$.
The solutions have very simple form, \begin{eqnarray}
\lambda_{0} & \approx & CG_{1}\frac{\mu}{T^{3}},\;\lambda_{1}\approx-CG_{2}\frac{\mu}{T^{4}},\;\lambda_{0}^{B}=\frac{\lambda_{0}}{\mu},\;\lambda_{1}^{B}=\frac{\lambda_{1}}{\mu}\label{eq:coeff-one-c}\end{eqnarray}
where we have used two constants, $G_{1}\equiv\frac{6750\pi^{2}\zeta(5)}{d_{g}G_{0}}$,
$G_{2}\equiv\frac{14}{d_{g}G_{0}}$. We notice that the $D$ terms
in Eq. (\ref{eq:sol-lambda01}) are negligible, so the solutions are
proportional to $\xi$. Note that the quantity $\frac{n\mu}{\epsilon+P}\sim\overline{\mu}^{2}$
is also small and we have dropped it, since $n\mu\approx\frac{d_{g}}{6\pi^{2}}T^{4}\overline{\mu}^{2}$
and $\epsilon+P=\frac{4}{3}\epsilon\approx d_{g}\frac{7\pi^{2}}{90}T^{4}$.

\section{With two charges and particle/antiparticle \label{sec:5}}

As an example for the case of two charges, we consider adding to the
system the chirality or an axial $U(1)$ charge to particles. Then
there are two currents, one for each chirality, or equivalently, for
the $U(1)/U_{A}(1)$ charge. For simplicity we assume that there is
an anomaly for the axial charge current but no anomaly for the charge
one. There are distribution functions for right-hand and left-hand
particles, $f_{a}^{Q}(x,p)$ ($a=R,L$), with chemical potentials
$\mu_{R.L}=\mu\pm\mu_{A}$. As an extension to Eq. (\ref{eq:deviation}),
the corrections $\chi_{a}(x,p)$ in $f_{a}^{Q}(x,p)$ are now $\chi_{a}(x,p)=\lambda_{a}p^{\mu}\omega_{\mu}+\lambda_{aB}p^{\mu}B_{\mu}$.
Instead of right-hand and left-hand quantities $X_{a}$, we can equivalently
use $X=X_{R}+X_{L}$ and $X_{A}=X_{R}-X_{L}$, where $X=\lambda,\lambda_{B},\xi,\xi_{B},n,s,\epsilon,P,j_{\mu}$.
The distribution functions $f_{a}^{Q}(x,p)$ satisfy two separate
Boltzmann equations of the following form, \begin{equation}
p^{\mu}\left(\frac{\partial}{\partial x^{\mu}}-QF_{\mu\nu}\frac{\partial}{\partial p_{\nu}}\right)f_{a}^{Q}(x,p)=\mathcal{C}_{aQ}[f_{b}^{Q'}].\label{eq:boltzmann-two-charge}\end{equation}
The $U(1)/U_{A}(1)$ charge and entropy currents and the stress tensor
in equilibrium are give by: $j_{0}^{\mu}=nu^{\mu}$, $j_{A0}^{\mu}=n_{A}u^{\mu}$,
$S_{0}^{\mu}=su^{\mu}$ and $T_{0}^{\mu\nu}=(\epsilon+P)u^{\mu}u^{\nu}-Pg^{\mu\nu}$,
with the particle number density $n\equiv\sum_{aQ}Qn_{aQ}$ ($a=R,L$;
$Q=\pm1$), the energy density $\epsilon=\sum_{aQ}\epsilon_{aQ}$,
the pressure $P=\sum_{aQ}P_{aQ}$, and the entropy density $s=\sum_{aQ}s_{aQ}=(\epsilon+P-\sum_{a}n_{a}\mu_{a})/T$. 

Similar to Eqs. (\ref{eq:djx},\ref{eq:constraint1}), the $U(1)/U_{A}(1)$
charge conservation equations (\ref{eq:charge-current}) can be derived
as \begin{eqnarray}
\partial_{\mu}j^{\mu}(x) & = & \int[dp](\mathcal{C}_{R,+}-\mathcal{C}_{R,+}+\mathcal{C}_{L,+}-\mathcal{C}_{L,+})=0,\label{eq:djx-1}\\
\partial_{\mu}j_{A}^{\mu}(x) & = & \int[dp](\mathcal{C}_{R,+}-\mathcal{C}_{R,-}-\mathcal{C}_{L,+}+\mathcal{C}_{L,+})=-CE^{\mu}B_{\mu},\label{eq:djx-2}\end{eqnarray}
where we have used Eq. (\ref{eq:property-1}). The energy and momentum
conservation equation reads, \begin{eqnarray}
\partial_{\mu}T^{\mu\nu}-F^{\nu\mu}j_{\mu} & = & \int[dp]p^{\nu}(\mathcal{C}_{R,+}+\mathcal{C}_{R,-}+\mathcal{C}_{L,+}+\mathcal{C}_{L,-})=0,\label{eq:em-two}\end{eqnarray}
where we have used Eq. (\ref{eq:property-2}). 

Similar to Eq. (\ref{eq:db}), we can express $\xi_{a}$, $\xi_{aB}$,
$D$ and $D_{B}$ in terms of $\lambda_{a}$ and $\lambda_{aB}$ as, 

\begin{eqnarray}
 &  & \xi_{a}=-(J_{21}^{\lambda_{a},+}-J_{21}^{\lambda_{a},-}),\;\;-DT=J_{31}^{\lambda_{R},+}+J_{31}^{\lambda_{R},-}+J_{31}^{\lambda_{L},+}+J_{31}^{\lambda_{L},-},\nonumber \\
 &  & \xi_{aB}=-(J_{21}^{\lambda_{aB},+}-J_{21}^{\lambda_{aB},-}),\;\;-D_{B}T=J_{31}^{\lambda_{RB},+}+J_{31}^{\lambda_{RB},-}+J_{31}^{\lambda_{LB},+}+J_{31}^{\lambda_{LB},-},\label{eq:xi-d}\end{eqnarray}
for $a=R,L$. We have the following first order corrections, \begin{eqnarray}
j_{1}^{\mu} & = & j_{R1}^{\mu}+j_{L1}^{\mu}=\sum_{a=R,L}(\xi_{a}\omega^{\mu}+\xi_{aB}B^{\mu})=\xi\omega^{\mu}+\xi_{B}B^{\mu},\nonumber \\
j_{A1}^{\mu} & = & j_{R1}^{\mu}-j_{L1}^{\mu}=(\xi_{R}-\xi_{L})\omega^{\mu}+(\xi_{RB}-\xi_{LB})B^{\mu}=\xi_{A}\omega^{\mu}+\xi_{AB}B^{\mu},\nonumber \\
T_{1}^{\mu\nu} & = & DT\left(u^{\mu}\omega^{\nu}+u^{\nu}\omega^{\mu}\right)+D_{B}T\left(u^{\mu}B^{\nu}+u^{\nu}B^{\mu}\right),\nonumber \\
S_{1}^{\mu} & = & -\sum_{a=R,L}\frac{\mu_{a}}{T}(\xi_{a}\omega^{\mu}+\xi_{aB}B^{\mu})+(D\omega^{\mu}+D_{B}B^{\mu})\nonumber \\
 & = & -\sum_{i=\mathrm{null},A}\frac{\mu_{i}}{T}(\xi_{i}\omega^{\mu}+\xi_{iB}B^{\mu})+(D\omega^{\mu}+D_{B}B^{\mu}).\label{eq:j-t-s2}\end{eqnarray}
It can be verified that the entropy current in Eq. (\ref{eq:j-t-s2})
cannot satisfy $\partial_{\mu}S^{\mu}\geq0$ unless $C=0$. In order
to ensure the positivity of $\partial_{\mu}S^{\mu}$ in presence of
an anomaly, one would have to substract the vector $Q^{\mu}=D\omega^{\mu}+D_{B}B^{\mu}$
from $S^{\mu}$. With $U(1)$ and $U_{A}(1)$ charges, we have \begin{eqnarray}
\widetilde{S}^{\mu} & = & S^{\mu}-Q^{\mu}=su^{\mu}-\sum_{a=R,L}\overline{\mu}_{a}(\xi_{a}\omega^{\mu}+\xi_{aB}B^{\mu})\nonumber \\
 & = & \frac{1}{T}(Pu^{\mu}-\sum_{a=R,L}\mu_{a}j_{a}^{\mu}+u_{\lambda}T^{\lambda\mu})-Q^{\mu}.\label{eq:entropy-current}\end{eqnarray}
In the same way as in Sec. \ref{sec:2}, the divergence of the entropy
current can be evaluated as, \begin{eqnarray}
\partial_{\mu}\widetilde{S}^{\mu} & = & \omega^{\mu}\left[\sum_{a=R,L}\partial_{\mu}\overline{\mu}_{a}\left(\frac{n_{i}TD}{\epsilon+P}-\xi_{a}\right)-\partial_{\mu}D+\frac{2D}{\epsilon+P}\partial_{\mu}P\right]\nonumber \\
 &  & +B^{\mu}\left[\sum_{a=R,L}\partial_{\mu}\overline{\mu}_{a}\left(\frac{n_{i}TD_{B}}{\epsilon+P}-\xi_{aB}\right)-\partial_{\mu}D_{B}+\frac{D_{B}}{\epsilon+P}\partial_{\mu}P\right]\nonumber \\
 &  & +E\cdot\omega\left[\sum_{a=R,L}\left(\frac{n_{a}D}{\epsilon+P}-\frac{\xi_{a}}{T}\right)+\frac{2nD}{\epsilon+P}-2D_{B}\right]\nonumber \\
 &  & +E\cdot B\left[\sum_{a=R,L}\left(\frac{n_{a}D_{B}}{\epsilon+P}-\frac{\xi_{aB}}{T}\right)+C\frac{\mu_{A}}{T}+\frac{nD_{B}}{\epsilon+P}\right],\label{eq:div-s}\end{eqnarray}
following Eqs. (\ref{eq:djx-1}-\eqref{eq:em-two}). By imposing all
quantities inside the square brackets to vanish we can solve $\xi_{a}$,
$\xi_{aB}$, $D$, $D_{B}$ as follows, \begin{eqnarray}
\xi & = & 2C\mu\mu_{A}\left(1-\frac{3}{2}\frac{n\mu}{\epsilon+P}\right),\;\xi_{A}=C\mu^{2}\left(1-\frac{3n_{A}\mu_{A}}{\epsilon+P}\right),\nonumber \\
\xi_{B} & = & C\mu_{A}\left(1-\frac{2n\mu}{\epsilon+P}\right),\;\xi_{AB}=C\mu\left(1-\frac{2n_{A}\mu_{A}}{\epsilon+P}\right),\nonumber \\
DT & = & -C\mu_{A}\mu^{2},\; D_{B}T=-C\mu_{A}\mu\end{eqnarray}
It is interesting to observe that for small $\mu$, $\mu_{A}$, we
obtain very simple and symmetric solutions, \begin{eqnarray}
 &  & \xi\approx2C\mu\mu_{A},\:\xi_{A}\approx C\mu^{2},\:\xi_{B}\approx C\mu_{A},\:\xi_{AB}\approx C\mu,\nonumber \\
 &  & DT=-C\mu_{A}\mu^{2},\; D_{B}T=-C\mu_{A}\mu.\label{eq:solution1}\end{eqnarray}
which is identical to the result of Ref. \cite{Fukushima:2008xe,Kharzeev:2010gr}.
Here we have assumed that all integral constants are vanishing. Note
that Eq. (\ref{eq:solution1}) is the result of the entropy principle
in the hydrodynamic approach which was also obtained in Ref. \cite{Sadofyev:2010pr,Amado:2011zx,Kalaydzhyan:2011vx}. 

Equivalently we can use $\xi_{R}=(\xi+\xi_{A})/2$ and $\xi_{L}=(\xi-\xi_{A})/2$
to determine $\lambda_{a}$, $\lambda_{aB}$ $(a=R,L)$ via solving
a system of equations (\ref{eq:xi-d}), where the first/second line
(each has three equations) is for $\lambda_{R,L}/\lambda_{RB,LB}$.
For minimal number of coefficients we can determine the values of
these coefficients completely, $\lambda_{R,L}$ and $\lambda_{RB,LB}$
can be expanded to the zeroth or first power of $u\cdot p$. For example,
if we expand $\lambda_{R}$ to the zeroth power, then we have to expand
$\lambda_{L}$ to the first power, and vice versa. Suppose we take
the former case, $\lambda_{R}=\lambda_{R0}$ and $\lambda_{L}=\lambda_{L0}+\lambda_{L1}(u\cdot p)$,
we can solve $\lambda_{R,L}$ as \begin{eqnarray}
\lambda_{R0} & = & -\xi_{R}\frac{1}{\delta J_{21}^{R}},\nonumber \\
\lambda_{L0} & = & -\frac{1}{\Delta}\xi_{L}(\sigma J_{41}^{L})+\frac{1}{\Delta}\left(DT-\xi_{R}\frac{\sigma J_{31}^{R}}{\delta J_{21}^{R}}\right)(\delta J_{31}^{L}),\nonumber \\
\lambda_{L1} & = & \frac{1}{\Delta}\xi_{L}(\sigma J_{31}^{L})-\frac{1}{\Delta}\left(DT-\xi_{R}\frac{\sigma J_{31}^{R}}{\delta J_{21}^{R}}\right)(\delta J_{21}^{L}),\end{eqnarray}
where $\Delta=(\delta J_{21}^{L})(\sigma J_{41}^{L})-(\delta J_{31}^{L})(\sigma J_{31}^{L})$.
The solutions to $\lambda_{RB,LB}$ take the same form as above with
replacements $\lambda_{a}\rightarrow\lambda_{aB}$, $\xi_{a}\rightarrow\xi_{aB}$
and $D\rightarrow D_{B}$ if we assume the same expansion as $\lambda_{R,L}$:
$\lambda_{RB}=\lambda_{RB,0}$ and $\lambda_{LB}=\lambda_{LB,0}+\lambda_{LB,1}(u\cdot p)$. 

For massless fermions and small $\overline{\mu}_{R,L}$ (or equivalently
small $\overline{\mu}$ and $\overline{\mu}_{A}$), we obtain \begin{eqnarray}
\sigma J_{21}^{R,L} & \approx & 9\zeta(3)G,\;\sigma J_{31}^{R,L}\approx\frac{7\pi^{4}}{15}GT,\;\sigma J_{41}^{R,L}\approx225\zeta(5)GT^{2},\nonumber \\
\delta J_{21}^{R,L} & \approx & \pi^{2}\overline{\mu}_{R,L}G,\;\delta J_{31}^{R,L}\approx36\zeta(3)\overline{\mu}_{R,L}GT,\;\delta J_{41}^{R,L}\approx\frac{7\pi^{4}}{3}\overline{\mu}_{R,L}GT^{2},\nonumber \\
\Delta & \approx & \frac{\pi^{2}}{5}G^{2}G_{0}T^{2}\overline{\mu}_{L},\end{eqnarray}
where $G$ and $G_{0}$ are the same as in the former section. Then
the solutions to $\lambda_{R,L}$ are \begin{eqnarray}
\lambda_{R0} & \approx & -\frac{3C}{d_{g}}\frac{\mu}{T^{3}}\frac{\mu+2\mu_{A}}{\mu+\mu_{A}},\;\lambda_{L0}\approx\frac{3C}{d_{g}}\frac{\mu}{T^{3}}\left(G_{4}\frac{\mu+2\mu_{A}}{\mu+\mu_{A}}+G_{5}\frac{\mu-2\mu_{A}}{\mu-\mu_{A}}\right),\nonumber \\
\lambda_{L1} & \approx & \frac{14\pi^{4}C}{d_{g}G_{0}}\frac{\mu}{T^{4}}\frac{\mu\mu_{A}}{\mu^{2}-\mu_{A}^{2}},\nonumber \\
\lambda_{RB,0} & \approx & -\frac{3C}{d_{g}}\frac{1}{T^{3}},\;\lambda_{LB,0}\approx\frac{3C}{d_{g}}\frac{1}{T^{3}},\;\lambda_{LB,1}\approx\frac{30\pi^{2}C}{d_{g}G_{0}}\frac{\mu\mu_{A}}{T^{6}},\end{eqnarray}
where $G_{4}\equiv-\frac{1}{G_{0}}84\pi^{2}\zeta(3)$ and $G_{5}\equiv\frac{1}{G_{0}}1125\zeta(5)$.
The coefficient ratios of $\lambda_{ai}/\lambda_{aB,i}$ ($i=0,1$)
are proportional to $\mu$ times dimensionless factors, \begin{eqnarray}
\frac{\lambda_{R0}}{\lambda_{RB,0}} & \approx & \mu\frac{\mu+2\mu_{A}}{\mu+\mu_{A}},\;\frac{\lambda_{L0}}{\lambda_{LB,0}}\approx\mu\left(G_{4}\frac{\mu+2\mu_{A}}{\mu+\mu_{A}}+G_{5}\frac{\mu-2\mu_{A}}{\mu-\mu_{A}}\right),\nonumber \\
\frac{\lambda_{L1}}{\lambda_{LB,1}} & \approx & \mu\frac{7}{15}\pi^{2}\frac{T^{2}}{\mu^{2}-\mu_{A}^{2}}.\end{eqnarray}

Note that $\lambda_{LB,1}\ll T\lambda_{LB,0}\sim T\lambda_{RB,0}$,
so both $\lambda_{BR}(p)$ and $\lambda_{BL}(p)$ can be constants
at small $\overline{\mu}$ and $\overline{\mu}_{A}$ limit. This property
is quite different from the one-charge case in which $\lambda_{B}(p)$
must have momentum dependence in order to comply with the entropy
principle.

\section{Discussions and conclusions \label{sec:6}}

We have shown that induced terms related to the vorticity and magnetic
field in the charge and entropy currents from a triangle anomaly can
be derived in kinetic theory by introducing correction terms to the
phase space distribution function at the first order. We demonstrated
that the anomalous source terms are necessary to ensure that the equations
for the charge and energy-momentum conservation are satisfied and
that the correction terms of distribution functions are compatible
to these equations. 

As examples for the correction terms of distribution functions, we
focus on the massless fermionic system in three cases for small $\mu/T$,
with one charge {[}$U(1)${]} and one particle species (without anti-particles),
with one charge {[}$U(1)${]} and particles/anti-particles, and with
two charges {[}$U(1)\times U_{A}(1)${]}. In the latter two cases,
the coefficients for $\omega$ and $B$ terms in distribution functions
are found to be proportional to $C\mu/T^{3}$ and $C/T^{3}$ respectively.
In the two-charges case the coefficients can be constants or independent
of momentum, such a property is impossible for the one-charge case
since it is not allowed by the entropy principle. 

In the two-charges case, we assumed that there is an anomaly for the
axial charge current but no anomaly for the charge one. The coefficients
of correction terms for the charge/axial-charge currents and energy-momentum
tensor have a very simple and symmetric form at small $\mu/T$ and
$\mu_{A}/T$ limit: $\xi\approx2C\mu\mu_{A}$, $\xi_{A}\approx C\mu^{2}$,
$\xi_{B}\approx C\mu_{A}$, $\xi_{AB}\approx C\mu$, $DT=-C\mu_{A}\mu^{2}$,
and $D_{B}T=-C\mu_{A}\mu$. This means that similar to the CME an
axial anomaly can induce a residual charge current which is proportional
to the magnetic field and the axial chemical potential \cite{Fukushima:2008xe}. 

We have a few comments about our results. In our evaluation of the
correction terms of distribution functions, we have assumed that $\lambda(p)$
and $\lambda_{B}(p)$ are identical to particles and anti-particles.
Alternatively we can assume that they have an opposite sign for particles
to anti-particles. We can not tell which case is correct due to lack
of deeper knowledge about these anomalous term at a microscopic level.
Similarly we have assumed that $\lambda(p)$ and $\lambda_{B}(p)$
have different values for right-handed particles from left-handed
ones. One can also assume that they have the same or opposite values
for right-handed and left-handed particles. In the current framework
one can not tell which is correct. Such a situation is like what happens
in an effective theory when many effective candidates point to a unique
microscopic theory. We also note that the solutions to $\lambda(p)$
and $\lambda_{B}(p)$ given in this paper are for the cases where
the number of unknown coefficients in $\lambda(p)$ and $\lambda_{B}(p)$
is equal to that of constraining equations. It is possible that $\lambda(p)$
and $\lambda_{B}(p)$ can be expanded to higher powers of $(p\cdot u)$
and then have larger number of unknown coefficients than that of constraining
equations. In this case the constraining equations just provide constraints
for $\lambda(p)$ and $\lambda_{B}(p)$ from the second law of thermodynamics. 

\textbf{Acknowledgement.} QW thanks J.W. Chen for suggesting this
work and for insightful discussions. QW is supported in part by the
{}``100 talents'' project of Chinese Academy of Sciences (CAS) and
by the National Natural Science Foundation of China (NSFC) with Grant
No. 10735040. JHG is supported in part by the China Postdoctoral Science
Foundation under Grant No. 20090460736.


\begin{thebibliography}{28}
\bibitem{Jacobs:2004qv}P.~Jacobs and X.~N.~Wang, Prog.\ Part.\ Nucl.\ Phys.\  {\bf 54}, 443 (2005)   [arXiv:hep-ph/0405125].   

\bibitem{Shuryak:2004cy}E.~V.~Shuryak, Nucl.\ Phys.\ A {\bf 750},
64 (2005) {[}arXiv:hep-ph/0405066{]}. 

\bibitem{gyulassy:2005}M.~Gyulassy, L.~McLerran, Nucl.\ Phys.\ A{\bf 750},
30(2005). 

\bibitem{Eckart:1940te}C.~Eckart, Phys.\ Rev.\  {\bf 58}, 919 (1940). 

\bibitem{landau:1959}L.\ D.\ Landau and E.\ M.\ Lifshitz, Fluid
Mechanics (Pergamon, New York, 1959). 

\bibitem{Israel:1976tn}W.~Israel, Annals Phys.\  {\bf 100}, 310 (1976). 

\bibitem{Israel:1979wp}W.~Israel and J.~M.~Stewart, Annals Phys.\  {\bf 118}, 341 (1979).   

\bibitem{Huovinen:2001cy}P.~Huovinen, P.~F.~Kolb, U.~W.~Heinz, P.~V.~Ruuskanen and S.~A.~Voloshin,   Phys.\ Lett.\  B {\bf 503}, 58 (2001)   [arXiv:hep-ph/0101136].   

\bibitem{Kolb:2003dz}P.~F.~Kolb and U.~W.~Heinz, arXiv:nucl-th/0305084. 

\bibitem{Song:2007fn}H.~Song and U.~W.~Heinz, Phys.\ Lett.\  B {\bf 658}, 279 (2008)   [arXiv:0709.0742 [nucl-th]].   

\bibitem{Muronga:2001zk}A.~Muronga, Phys.\ Rev.\ Lett.\  {\bf 88}, 062302 (2002)   [Erratum-ibid.\  {\bf 89}, 159901 (2002)]   [arXiv:nucl-th/0104064].   

\bibitem{Muronga:2004sf}A.~Muronga and D.~H.~Rischke, arXiv:nucl-th/0407114.

\bibitem{Romatschke:2007mq}P.~Romatschke and U.~Romatschke, Phys.\ Rev.\ Lett.\  {\bf 99}, 172301 (2007)   [arXiv:0706.1522 [nucl-th]].   

\bibitem{Baier:2007ix}R.~Baier, P.~Romatschke, D.~T.~Son, A.~O.~Starinets and M.~A.~Stephanov,   JHEP {\bf 0804}, 100 (2008)   [arXiv:0712.2451 [hep-th]]. 

\bibitem{Betz:2008me}B.~Betz, D.~Henkel and D.~H.~Rischke, Prog.\ Part.\ Nucl.\ Phys.\  {\bf 62}, 556 (2009)   [arXiv:0812.1440 [nucl-th]].   

\bibitem{Erdmenger:2008rm}J.~Erdmenger, M.~Haack, M.~Kaminski and A.~Yarom,   
JHEP {\bf 0901}, 055 (2009)   [arXiv:0809.2488 [hep-th]].   

\bibitem{Banerjee:2008th}N.~Banerjee, J.~Bhattacharya, S.~Bhattacharyya, 
S.~Dutta, R.~Loganayagam and P.~Surowka,   
arXiv:0809.2596 [hep-th].   

\bibitem{Torabian:2009qk}M.~Torabian and H.~U.~Yee,   
JHEP {\bf 0908} (2009) 020   [arXiv:0903.4894 [hep-th]].   

\bibitem{Rebhan:2009vc}A.~Rebhan, A.~Schmitt and S.~A.~Stricker,   
JHEP {\bf 1001}, 026 (2010)   [arXiv:0909.4782 [hep-th]].   

\bibitem{Son:2009tf}D.~T.~Son and P.~Surowka, Phys.\ Rev.\ Lett.\  {\bf 103}, 191601 (2009)   [arXiv:0906.5044 [hep-th]].   

\bibitem{Lublinsky:2009wr}M.~Lublinsky and I.~Zahed,   
Phys.\ Lett.\  B {\bf 684}, 119 (2010) [arXiv:0910.1373 [hep-th]].   

\bibitem{Kharzeev:2007jp}D.~E.~Kharzeev, L.~D.~McLerran and H.~J.~Warringa, Nucl.\ Phys.\  A {\bf 803}, 227 (2008)   [arXiv:0711.0950 [hep-ph]].   

\bibitem{Fukushima:2008xe}K.~Fukushima, D.~E.~Kharzeev and H.~J.~Warringa, Phys.\ Rev.\  D {\bf 78}, 074033 (2008)   [arXiv:0808.3382 [hep-ph]].   

\bibitem{Fukushima:2010vw}K.~Fukushima, D.~E.~Kharzeev and H.~J.~Warringa, Phys.\ Rev.\ Lett.\  {\bf 104}, 212001 (2010)   [arXiv:1002.2495 [hep-ph]].   

\bibitem{Kharzeev:2010gr}D.~E.~Kharzeev and D.~T.~Son, arXiv:1010.0038 [hep-ph].   

\bibitem{Sadofyev:2010pr}A.~V.~Sadofyev and M.~V.~Isachenkov, Phys.\ Lett.\  B {\bf 697}, 404 (2011)   [arXiv:1010.1550 [hep-th]].   

\bibitem{Amado:2011zx}I.~Amado, K.~Landsteiner and F.~Pena-Benitez, arXiv:1102.4577 [hep-th].   

\bibitem{Kalaydzhyan:2011vx}T.~Kalaydzhyan and I.~Kirsch,   arXiv:1102.4334 [hep-th].   
\end{thebibliography}
\end{document}